# STRUCTURAL PHASE TRANSITIONS AND SOUND VELOCITY ANOMALIES IN $La_{1-x} Sr_x MnO_3$

Yu.P.Gaidukov[a,1], N.P.Danilova[a], N.A.Vassilieva[a], A.M.Balbashov[b]

[a] *Department of Physics, M. V. Lomonosov Moscow State University, 119899 Moscow, Russia*
[b] *Moscow Power Engineering Institute, 105835 Moscow, Russia*

The sound velocities dependencies on temperature V(T) were measured by means of resonance method have on the thin plates of quasimonocrystalline manganites $La_{1-x}Sr_xMnO_3$. Results are presented for x=0.1, 0.125, 0.15, 0.175 and 0.2 in the temperature range from 70 to 350K. The V(T) behavior correlates with those of susceptibilities $\chi_{ac}$(T) and magnetic moment M(T). For x=0.175 and 0.2 the transition to ferromagnetic state is accompanied by appearance of large but smooth rise in V(T) whereas structural phase transition between the rhomboedral (R) and orthorhombic (O) structures is accompanied by sharp increase up to 15-20% in V(T). For x=0.1, 0.125 and 0.15 the structural phase transition between Jahn-Teller distorted orthorhombic structure (O') and pseudocubic one (O'') is accompanied by deep minimum (≈20-30%) in V(T).

*PACS*: 62.65.+k; 75.30.Kz; 64.70.Rb; 43.20.Hq

*Keywords*: manganites $La_{1-x}Sr_xMnO_3$; ultrasound; sound velocity; structural phase transition; susceptibility.

## 1. Introduction

Below there are presented the more detailed than previously [1] results of sound velocities' investigations in perovskite-type manganite $La_{1-x} Sr_x MnO_3$ over the wide temperature range. The goal of these investigations was to establish correlation between elastic moduli and various magnetic and structure transitions.

## 2. Experimental method

The plate-like samples of different shape (discs, squares, rectangles and half-disks) were cut from quazimonocrystalls prepared by floating zone melting with radiation heating [2]. The thickness of samples were ≈1 mm, the other dimensions were of 2.5 to 8 mm. The sound velocities were determined from resonance acoustic frequencies [3] observed in the frequency range f=0.1-3 MHz in magnetic fields H=500 Oe. Isotropic model has been used to obtain numerical values for sound velocities from resonance frequencies. The excitation of elastic vibrations in the sample was caused by contact-free method through direct and inverse magnetostrictions [1,3]. In additional to these measurements the ac magnetic susceptibilities $\chi_{ac}$ were examined on f=0.3MHz. The magnetic field dependence of magnetic moment $M_s$(H) were obtained by induction magnetometer at given temperatures. The temperature range of measurements was T=70-350K.

## 3. Results and discussion

---
[1] Corresponding author: Fax:+7(095) 932 88 76; e-mail: gaidukov@lt.phys.msu.su

Fig.1 presents the typical behavior of the transverse sound velocities $V_t$ as a function of T for x=0.175 and 0.2 (here and further all the dependencies were taken on temperature increase).

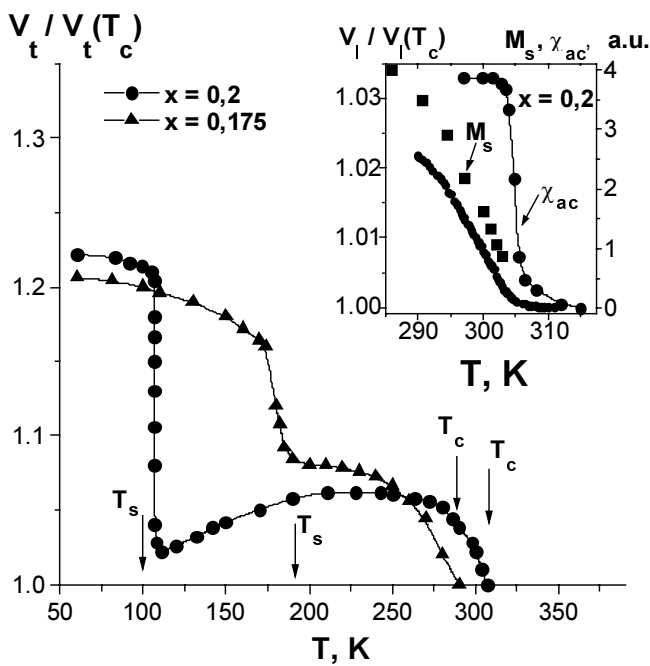

Fig. 1.

The vertical arrows indicate the magnetic (paramagnetic-ferromagnetic, $T_c$) and structural (R→O, $T_s$) phase transitions according to [4]. From this figure one can see that the sound velocity depends only slightly on the temperature (see inset on fig.1). The sudden pronounced rise in V(T) by 6-8% occurs below $T_c$. These V(T) behavior are attributed to influence of $La_{0.8}Sr_{0.2}MnO_3$ magnetic state on elastic moduli of samples. This is confirmed by correlation in behaviors of V(T) and $M_s(T)$. Far from $T_c$ another mechanism arouses caused softening of elastic moduli. As a consequence V(T) has attained a gentle maximum or show a decrease. At $T_s$ the sound velocity increases abruptly by 10-20% and here a temperature hysteresis (≈10-15K) has been observed. This indicates the first order of the structural transition. As the experiments show in the vicinity of $T_c$ a spontaneous magnetization $M_s$ undergo jump-like decrease of ≈4% (in the vicinity of 0.15 μB on Mn) and susceptibility - of ≈25%. Simultaneously with this jump-like decrease amplitudes of acoustic resonance increase several times. This is evidence that this new state is characterized by increased magnetostiction [5].

For x=0.15 V(T) and $\chi_{ac}(T)$ dependencies are presented on Fig.2. As temperature decrease the sound velocities decrease significantly, attaining minimum at $T_{min}$ ≈205K. A strong increase of one occurs with further temperature decrease. For one of the samples of this composition it has been possible to determine the transverse velocity $V_t$, longitudinal three-dimensional one $V_l$ and the velocity of zeroth longitudinal symmetric Lamb wave $V_l^{plate}$. At $T_{min}$ these ones are equal respectively to $V_t$ =2.3, $V_l$ =5.7 and $V_l^{plate}$ =3.9 (in units of $10^5$cm/sec). These values lead us to Poisson's ratio σ =0.4. As the temperature decreases further the Poisson's ratio decreases too and at T=150K σ =0.35. Special attention must be given to the fact that in the vicinity of $T_{min}$ there has been not observed any peculiarity in $\chi_{ac}(T)$ (H=20Oe) behavior. Nevertheless we are inclined to believe that $T_{min}$ =205K corresponds to structural phase transition between Jahn-Teller distorted orthorhombic structure (O') and pseudocubic one (O'') (the temperature hysteresis ≈5° also occurs here). A discrepancy between this assumption and diagram from [4] (see vertical arrows on fig.2) is attributable to probable discrepancy between stoihiometric composition of sample and nominal one. In any case according to behavior of $\chi_{ac}(T)$ the magnetic transition occurs above $T_{min}$ at T≈220K. It is conceivable that appreciable softening of elasticity below TO'→O'' and maximum in $\chi_{ac}(T)$ (H=20 Oe) are not caused by the same reason. Under H=500 Oe the maximum in $\chi_{ac}(T)$ shifts to a higher temperature and a sudden sharp jump with the following break appeared on the $\chi_{ac}(T)$ curve at T≈220K. The position of the latter is not varied under further increase of the magnetic field. On the fig.2 the $\chi_{ac}(T)$ curves for H=20 Oe and 500 Oe are shown at the different scales: the scale for $\chi_{ac}(T)$ (H=500 Oe) is four times larger than one for H=20 Oe.

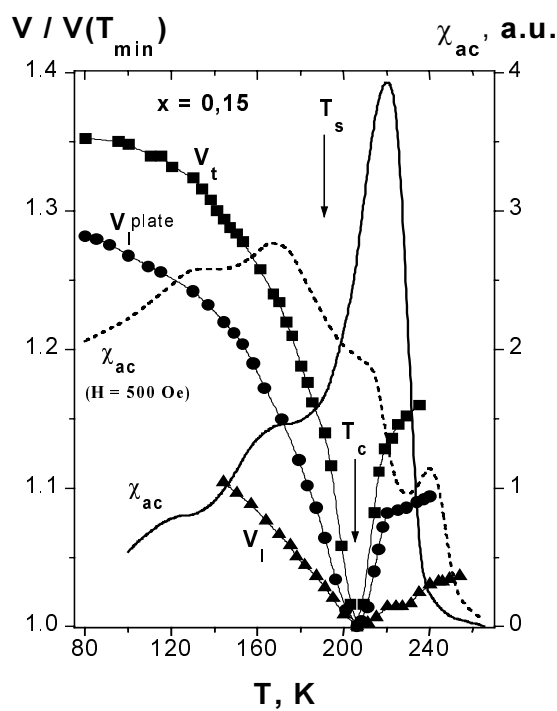

Fig. 2.

The phase diagram in the vicinity of x=0.15 is too complicated to be determined unambiguously from our results.

For x=0.125 the temperature dependence of $\chi_{ac}(T)$ (Fig.3) might be interpreted as follows. In the vicinity of T=155 K there is a magnetic transition between paramagnetic phase and canted antiferromagnetic (CA) one which is followed by ferromagnetic phase at T≈138K. Some difference between these temperatures and ones indicated in [4] (see the arrows on fig.3) are attributed to a probable less value of x (x=0.115 will suffice for these). As temperature decreases the $V_t(T)$- dependence is characterized by sharp decreasing of sound velocity, deep minimum at T=138K, further jump up to 15% and then by gradual increasing some more than 15%. Minimal value is $V_t≈1,9·10^5$ cm/sec. The reason of gradual increasing of $V_t(T)$ should be magnetic state of sample whereas the jump of $V_t(T)$ at T=138K should be attributed by the structural phase transition between Jahn-Teller distorted orthorhombic (O') phase and pseudocubic (O'') one (here, as well as for x=0.15, the temperature hysteresis ≈5° is observed). We failed to find any other transitions.

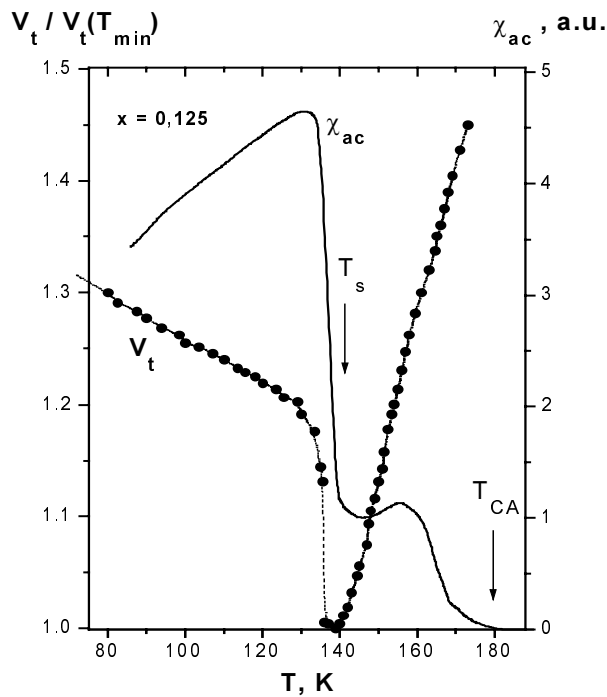

Fig. 3.

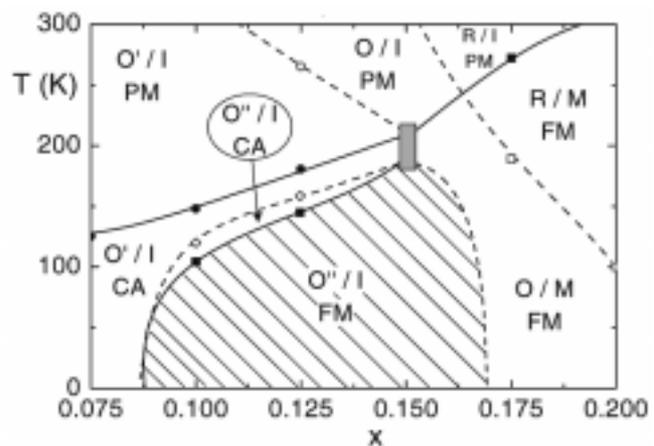

**$La_{1-x}Sr_xMnO_3$ phase diagram.** [4].

We are going to emphasize that magnetic field dependencies of sound velocities V(H) demonstrated that at $T>T_{min}$ the velocities decrease as H increase (about 5-7% at $H=10^4$ Oe) while at $T<T_{min}$ - they do increase slightly. As it is has been known that the magnetic field influence on elastic moduli in antiferromagnetic state is greater than in ferromagnetic one, it might additionally testify that we are dealing with antiferromagnetism above $T_{min}$ and with ferromagnetism below $T_{min}$.

The dependencies $\chi_{ac}(T)$ and $V_t(T)$ for x=0.1 reproduce ones for x=0.125 with a somewhat different value of peculiar temperatures: $T_{CA}=150K$, $T_{O'\rightarrow O''}=130K$

Hence the results of our investigations indicate that for x<0.175 the structural transitions O'→O'' and CA→FM coincide. For all the contents with x≤0.2 a strong correlation occurs between magnetic state of compounds and the temperature dependencies of elastic moduli.

## Acknowledgements

We are grateful to Ya.M.Mukovskiiy who provided us with a high-quality sample. We are also grateful to R.Z.Levitin for valuable consultations. The work was performed under the State Program for Physics of Quantum and Wave processes and supported by the Russian Foundation for Basic Research (Projects no.98-02-17401 and no.97-02-17325).